\documentclass{emulateapj}
\slugcomment{{\em}}
\usepackage{epsf}

\begin{document}
\title{A simple model for the size-evolution of elliptical galaxies}

\author{Sadegh Khochfar \&  Joseph Silk}
\affil{ Department of Physics, Denys Wilkinson Bldg., University
of Oxford, Keble Road, Oxford, OX1 3RH, UK }
\email{sadeghk@astro.ox.ac.uk; silk@astro.ox.ac.uk}

\begin{abstract}
We use semi-analytical modelling of galaxy formation to predict the 
redshift-size-evolution of elliptical galaxies. Using a simple 
model in which relative sizes of elliptical galaxies of a given mass 
correlate with the fraction of stars  formed in a star burst 
during a major merger event, we are  able to reproduce the observed 
redshift-size-evolution. 
The size evolution is a result of the amount of cold gas available 
during the major merger. 
Mergers at high redshifts are gas-rich and produce ellipticals with smaller 
sizes. In particular we find a power-law relation between the sizes at 
different redshifts, with the power-law index  giving a measure of the relative amount of
dissipation during the mergers that lead to the formation of an elliptical.
The size evolution is found to be stronger for more massive galaxies 
as they involve more gas at high redshifts when they form, compared to less 
massive ellipticals. Local ellipticals more massive than $5 \times 10^{11}$ 
M$_{\odot}$ will be approximately 4 times larger than their counterparts at 
$z=2$. Our results indicate that the scatter in the size of 
similar massive present day elliptical galaxies is a result of their 
formation epoch, with smaller ellipticals being formed earlier.
\end{abstract}


\section{Introduction}
Elliptical galaxies have been shown to have formed the bulk of their stars at high
redshifts within an intensive star burst \citep[e.g.][]{c04,t05}, supporting
 a scenario in which elliptical galaxies formed in a 'monolithic' collapse 
and then continued evolving passively. On the other hand, many stellar 
dynamical properties of elliptical galaxies can be explained by the merger of 
 similar massive galaxies (e.g. Barnes \& Hernquist 1992; Naab \& Burkert 
2003; Naab, Jesseit \& Burkert 2006) as initially proposed by \citet{tt72}. 
The predicted merger rate of galaxies within the CDM paradigm and the 
observations are in fair agreement \citep{kb01}. As to the 
mainly old stellar populations of elliptical galaxies, \citet{de06} show that  
these can be recovered within the merger scenario  which comes as a consequence
of massive elliptical galaxies forming by dry mergers 
between elliptical galaxies \citep{kb03}. In addition, 
it has been shown that dry mergers can indeed  explain 
the kinematical properties of massive elliptical galaxies 
(Khochfar \& Burkert 2005; Naab, Khochfar \& Burkert 2006). 
and might 
be able to even account for the partially depleted cores of luminous 
ellipticals \citep[e.g.][]{g04}.

The size evolution of elliptical galaxies could serve
as an additional possible test for these two models.
Recent size measurements of ellipticals carried out at high redshift indicate 
that they are much smaller than their local counter parts \citep{da05,tru05}. 
Numerical simulations by \citet{nt05} which use self-consistent cosmological
orbital parameters \citep{kb04} to set up mergers between disk galaxies,
find that the size of the remnant is of the same order as the size of the 
progenitor disk. The simulations of \citet{nt05} did not include 
any gas or star 
formation in contrast to simulations carried out by \citet{sh05}. The latter 
authors find that those stars which existed in the progenitor 
disks before the merger, which we call the {\it quiescent} component, 
have a $\sim 5.7$ times larger effective radius than the stars 
which formed during the merger in a violent star burst, which we will 
call the {\it merger} component of the elliptical. However, the ratio of the 
effective radii of the two components is very likely to depend on the 
structure of the progenitors and on the way the merger takes place. The 
total effective radius of the remnant will depend on the mass fraction of
each component, with a larger merger component leading to a smaller remnant.

Recently, \citet[][hereafter KS]{ks05} investigated the fraction of merger 
and quiescent components in early-type galaxies, finding that the quiescent 
component is decreasing with redshift and increasing with mass up
to a  characteristic mass scale $M_C=3 \times 10^{10} M_{\odot}$ 
\citep[][hereafter K03]{k03} at which it becomes constant.
In this letter we follow up on their study and predict the size-evolution of
elliptical galaxies within the CDM-paradigm.

\section{Model}
We use semi-analytical modelling of galaxy formation to predict the merger 
and quiescent components of elliptical galaxies. The dark matter history is 
calculated using the merger tree proposed by \citet{som99} with a mass 
resolution of $2 \times 10^9 M_{\odot}$. The baryonic 
physics within these dark matter halos is calculated following recipes 
presented in \citet{spr01} including a model for the reionizing background 
by \citet{som02}. In our simulation, we assume that elliptical galaxies 
form whenever a major merger ($M_1 /M_2 \leq 3.5$ with $M_1 \geq M_2$) takes 
place. We assume that during this process all the cold gas which was in the
 progenitor disks will be consumed in  a central starburst, adding to the 
spheroid mass, and that all stars in the progenitor disks will be 
scattered into the spheroid too. Furthermore we allow the stars of satellite
 galaxies in minor mergers to also contribute to the spheroid.
 During the evolution of a galaxy, we keep track of the origins of all stars 
brought into the spheroid and attribute them to two categories, merger and 
quiescent, where the first incorporates stars formed during a starburst 
in a major merger and the latter includes stars previously formed in a disk 
and added to the spheroid during a major merger. Each star will 
carry along its 
label and not change it, which means that if a star was made 
in a merger of two progenitor galaxies and the remnant of that merger 
participated in another merger, the star will  still contribute to the 
merger component of the final remnant.

  For more modelling details, we refer the 
reader to KS and references therein.
 Throughout this paper, we use the following set of cosmological parameters:
$\Omega_0=0.3$, $\Omega_{\Lambda}=0.7$, $\Omega_b/\Omega_0=0.15$, 
$\sigma_8=0.9$ and $h=0.65$. We note here that running our code with the 
latest cosmological parameters from the WMAP mission 
\citep{2006astro.ph..3449S} 
 only changes our results slightly and that most of the effects are 
compensated for by the free model parameters for the star formation 
efficiency and the supernova feedback.

\section{Relative Sizes of Ellipticals}
As we have mentioned above, the relative effective radii of merger and 
quiescent components are very likely to be dependent on the 
physical properties under which the merger takes place. To minimise 
this effect,  we compare the 
relative sizes for ellipticals of similar mass. 
By doing so, we can assume that 
e.g. the potential depth is the same and that feedback effects have 
the same efficiency. Furthermore, it has been shown that the orbital 
parameters of merging halos are on average independent of the remnant 
halo mass  and redshift \citep{kb04}, which allows us to assume that 
merging will take place on average with the same orbital set-up.
\begin{figure}
  \plotone{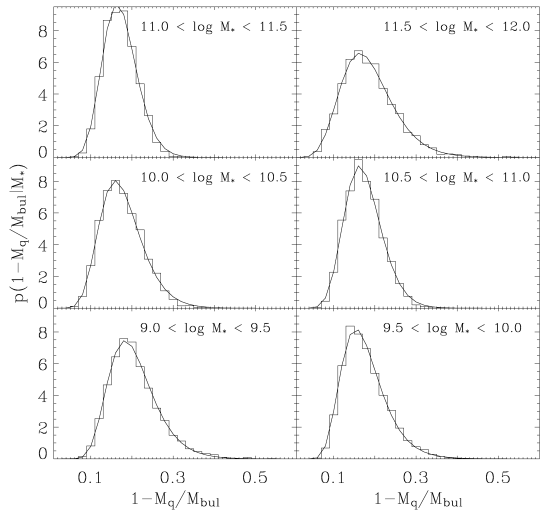}
    \figcaption{Histograms show the conditional probability density  
      $p(1-M_{q}/M_{bul}|M_*)$ of the merger component 
      in spheroids as a function of stellar mass 
      $M_*$  in six different mass bins. Solid lines show log-normal 
      distributions fitted to the data. Results are shown for elliptical      
      galaxies only.
      \label{fig6}}  
\end{figure}

The SDSS study revealed that the size distribution of galaxies with the 
same mass is log-normal and that the variance $\sigma_{\ln R_{\mbox{e}}}$ 
of the distribution is a function  of mass \citep[K03;][hereafter S03]{sh03}. 
Above $\sim 10^{10}$ M$_{\odot}$, the variance drops until it becomes 
approximately constant for galaxies more massive than
$\sim 10^{11}$ M$_{\odot}$.
 S03  show that their model, in which 
elliptical galaxies form by continued merging of galaxies, could reproduce the 
scatter in the size distribution of massive elliptical galaxies, by assuming 
the size distribution of the progenitors to be log-normal. 
However, the origin of the scatter in the size distribution of the 
progenitors remains not well understood. 
We start off by investigating the scatter in the merger
component and comparing it to the  scatter in the observed size distribution 
of early-type galaxies.
\begin{figure}
  \plotone{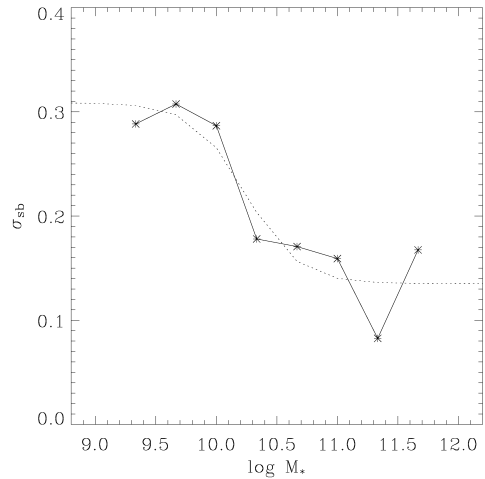}
    \figcaption{Dependence of the variance $\sigma_{\mbox{sb}}$ 
      of the conditional
      probability density $p(1-M_{q}/M_{bul}|M_*)$ as a function 
      of stellar mass $M_*$ of elliptical galaxies. The dotted line shows the
      fit using the empirical formula of S03.\label{fig7}}  
\end{figure}

In Fig. \ref{fig6}, we show the distribution of the 
merger components found in elliptical galaxies of various masses $M_*$,
 where elliptical galaxies are those having more than $65 \%$ of their 
stellar mass in their bulge component. We have
fitted the data with log-normal distributions following K03 and S03 as:
\begin{equation}
  p\left(x|M_*\right)=\frac{1}{\sqrt{2 \pi}(x-a)\sigma_{\mbox{sb}}} 
  \exp\left(-\frac{\ln^2[(x-a)/b]}{2 \sigma_{\mbox{sb}}^2}  \right)
\end{equation}
with $a$, $b$ and $\sigma_{\mbox{sb}}$ as free parameters, 
$x \equiv 1-M_q/M_{bul}$ as the merger component in a spheroid of mass 
$M_{bul}$, and $M_q$ as the quiescent component of that spheroid.
As can be seen, the log-normal distribution provides excellent fits to 
our simulated data. We can now try to compare  our variance 
$\sigma_{\mbox{sb}}$ as a function of galaxy mass to 
the observed variance $\sigma_{\ln R_{\mbox{e}}}$  in the SDSS.
S03 give an empirical fitting formulae to their results, 
\begin{equation}\label{eq6}
 \sigma_{\mbox{sb}}=\sigma_2 + \frac{\sigma_1-\sigma_2}{1+(M_*/M_0)^2}
\end{equation}
with $\sigma_1$, $\sigma_2$ and $M_0$ as free parameters. The fit of Eq. 
\ref{eq6} to our data is shown in Fig. \ref{fig7}. Our data seems to be fitted 
well by Eq. \ref{eq6} and the  trend in the observations can be 
recovered. The mass scale $M_0$, characteristic for 
 the transition point from large 
scatter to small scatter in the distributions is best fitted by a  value of 
$M_0\approx 1.7 \times 10^{10}$ M$_{\odot}$ in  our simulations, 
which is a about a factor 2 smaller than the 
value of $3.89 \times 10^{10}$ M$_{\odot}$ suggested by S03 but still
in reasonable agreement. For completeness we also give the values of the other 
fitting  parameters which are $\sigma_1=0.31$ and $\sigma_2=0.14$.

The results presented here suggest that the scatter in merger components 
$1-M_q/M_{bul}$ behaves 
like  the scatter in the size distribution of elliptical galaxies. We now 
make the simplified assumption that for each mass bin 
\begin{equation}\label{sca}
  \ln(1-M_{q}/M_{bul}) \propto \ln(R_{\mbox{e}}),
\end{equation}
where $R_{\mbox{e}}$ is in units of kpc and the proportionality constant 
reflects the average physical 
conditions that led to the formation of the elliptical galaxy and.
We note that other effects may influence the relative 
sizes of elliptical galaxies too
and that we here only focus on the contribution to it by the merger component. 
Using the assumption of proportionality one can now 
calculate the relative sizes of elliptical galaxies of the same mass by 
knowing their different merger components and using:
\begin{equation}\label{resca}
R_{\mbox{e}}(z_1)=R_{\mbox{e}}^{1/d}(z_0) \quad \mbox{with  } z_1 > z_0,
\end{equation}
with $d \equiv \ln (1-M_{q,0}/M_{bul})/\ln (1-M_{q,1}/M_{bul})$ as the 
dissipation factor which gives a measure of the relative amount of dissipation
that led to the formation of an elliptical.
The scatter in the size distribution of ellipticals decreases 
with mass and later becomes constant. The reason for the same behaviour in the 
scatter of the merger component is that  most massive ellipticals 
have their last major merger in a small redshift window not too 
far back in time. As a consequence the conditions regarding the gas fraction
involved in the merger are very similar and the scatter is small. 
\begin{figure}
  \plotone{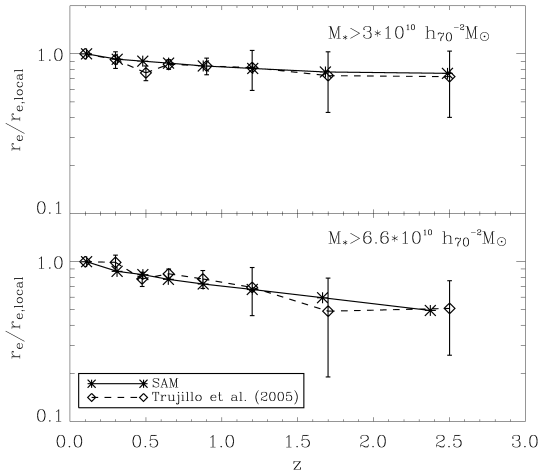}
    \figcaption{The evolution of sizes for early-type galaxies with respect 
      to the sizes of their local counter parts at a redshift $z=0.1$. 
      The upper panel shows the size-evolution   for early-type galaxies
      larger than $3 \times 10^{10} h_{70}^{-2}$ M$_{\odot}$ and
      the lower panel for early-type galaxies
      larger than $6.6 \times 10^{10} h_{70}^{-2}$ M$_{\odot}$.
      \label{fig11}}  
\end{figure}

\section{Size Evolution of Ellipticals}
\begin{figure}
  \plotone{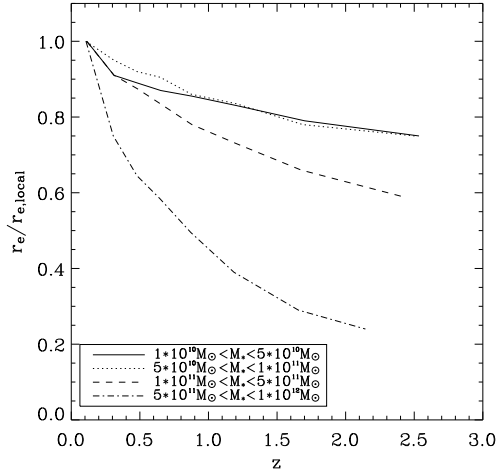}
    \figcaption{The predicted evolution of sizes for early-type 
      galaxies with respect 
      to the sizes of their local counter parts at a redshift $z=0.1$ divided
      into four different mass bins.
      \label{fig12}}  
\end{figure}

We now can test size evolution as a function of redshift
 predicted by Eq. \ref{resca} 
for elliptical galaxies of a given mass and compare it to the observations. 
Since we can only predict relative sizes between elliptical galaxies of  
approximately the same mass, we will normalise sizes to the SDSS sample. 
We calculate the size evolution for the same redshifts presented in 
\citet{tru05}. The authors took the mean effective radii 
of the $\ln(R_{\mbox{e}})$ 
distribution for galaxies above two mass thresholds of 
$3 \times 10^{10} h_{70}^{-2}$ M$_{\odot}$
and $6.6 \times 10^{10} h_{70}^{-2}$ M$_{\odot}$ from the SDDS sample of 
early-type galaxies and divided the effective radii of early-type galaxies at 
higher redshifts by this value. After arranging their galaxies in various 
redshift bins they calculated the means of these ratios and presented 
these values. We here use the same method to compare our results to theirs. Our
zero-point for individual galaxies is taken to be the mean value of 
$1-M_q/M_{bul}$ in our 
$\ln(1-M_q/M_{bul})$ distribution for galaxies of the same mass 
at a redshift of $z=0.1$. In Fig. \ref{fig11}
we show the expected evolution of sizes. For both cases of limiting 
masses, the agreement is excellent. It appears that the difference in sizes is 
more significant for massive early type galaxies. In Fig. \ref{fig12} we 
predict the  size-evolution in four different mass ranges based on the 
relative amount of their merger component. While local early-type
galaxies between $10^{10}$ M$_{\odot}$ and $10^{11}$ M$_{\odot}$ are around 
1.25 times larger than their counterparts at $z=2$, 
local  early types with masses larger than $5 \times 10^{11}$ 
M$_{\odot}$ will be approximately 4 times larger than their counterparts at 
$z=2$.  
This dramatic change in sizes in our model results from 
massive galaxies at high redshifts forming 
in  gas-rich mergers  while 
galaxies of the same mass at low redshifts form from gas-poor mergers (KS). 
This size-evolution might be an overestimate as the modelled galaxies suffer 
from over cooling of gas (KS) which is likely to overestimates the merger 
component at high redshift and the quiescent component at low redshift due 
to the shorter time between consecutive major mergers at high redshifts 
compared to low redshifts. 

\section{Discussion and Conclusions}
We have connected the scatter in the merger components $1-M_q/M_{bul}$ with 
the scatter in the sizes of elliptical galaxies. Following KS, the scatter in
the merger components of elliptical galaxies is a result of different 
formation epochs. Ellipticals forming early have  larger merger components, as 
they were formed in gas-rich mergers, hence  ellipticals with smaller 
effective radii must have formed earlier. We tested this assumption 
by predicting the size-evolution of elliptical galaxies at different 
redshifts and the agreement with the 
data of \citet{tru05} is excellent. It is important to note that we 
normalise the proportionality in our relation between the merger component 
and the effective radius by the observed scatter in the local galaxy sample, 
and then  go ahead and predict how sizes at earlier
redshifts  compare to  local sizes. 

Our results demonstrate  that the strongest size evolution is for  massive elliptical galaxies.
Local  early-types with masses larger than $5 \times 10^{11}$ 
M$_{\odot}$ will be approximately 4 times larger than their counterparts at 
$z=2$. This extreme size-evolution is a reflection of the progenitors having 
larger gas fractions at high redshifts and becoming more 'dry' at low 
redshifts.
In addition the low redshift progenitors have stellar disks that are more 
massive than the available amount of cold gas, hence increasing the quiescent 
fraction of the remnant. The progenitors however, are bulge dominated with
a disk component not more massive than $\sim 20\%$ of the total mass, thus the
size-evolution appears to be connected to the occurrence of bulge-dominated
dry mergers with time. The most massive elliptical galaxies in our simulations
undergo on average between one and two substantial dry mergers between 
$z=2$ and today.

We find 
that the relative amount of dissipation involved in the mergers relates to 
the size by a power-law as described in Eq. \ref{resca} where  the power-law 
exponent $d$ is the dissipation factor.
If this relation holds one can try to measure the relative amount of 
dissipation by measuring the relative sizes of recently formed ellipticals 
of the same mass at different redshifts.
Even though theoretically it sounds  straightforward to measure the 
dissipation factor we acknowledge that it is observationally not an easy task.
 The main problem here will be to identify elliptical galaxies that 
just formed. This is important as e.g. continued accretion of satellites 
or cold gas and subsequent star formation will alter the size of the 
merger remnant. One way of identifying recently formed ellipticals at high 
redshift might be by looking for signs of recent star formation. If 
there is sufficient gas involved in the major merger, 
simulations show that a starburst will be ignited \citep[e.g.][]{ba92} whose 
signature may be measurable. 

Our results presented here support a picture in which ellipticals form 
in mergers.
Future observations of high redshift ellipticals will allow one to make more 
accurate comparisons to the model we introduced here and will allow us to 
estimate the role of dissipational processes during major mergers of galaxies.
\acknowledgments
We would like to thank Ignacio Trujillo for providing the data in Fig. 
\ref{fig11} and for his helpful comments and for the comments of the referee 
which helped improving the paper.

\end{document}